\input{aipcheck.tex}

\documentclass[final]  {aipproc}
\layoutstyle{6x9}

\usepackage{amsmath}
\DeclareMathOperator{\tr}{tr}
\DeclareMathOperator{\Tr}{Tr}
\usepackage{calc}
\begin{document}

\title
      {Relativistic Faddeev approach to a non-local NJL model}

\author{Amir H. Rezaeian}{
address={Department of Physics,UMIST, PO Box 88, Manchester, M60 1QD, UK},
email={Rezaeian@Theory.phy.umist.ac.uk}
}
\author{Niels R. Walet}{
address={Department of Physics,UMIST, PO Box 88, Manchester, M60 1QD, UK},
email={Niels.Walet@Umist.ac.uk}
}
\author{Michael C. Birse}{
address={The Theoretical Physics Group, Department of Physics and Astronomy, University of
Manchester, Manchester, M13 9PL, UK},
email={Mike.Birse@man.ac.uk}
 }

\copyrightyear  {2003}

\begin{abstract}
The diquark and nucleon are studied in a non-local NJL model. We solve
the relativistic Faddeev equation and compare the results with the
ordinary NJL model. Although the model is quark confining, it is not
diquark confining in the rainbow-ladder approximation. We show that
the off-shell contribution to the diquark $T$ matrix is crucial for
the structure of the nucleon: without its inclusion the attraction in
the scalar channel is too weak to form a three-body bound state.
\end{abstract}

\date{\today}

\maketitle

\section{Introduction}
The NJL model is a successful low-energy phenomenological
model inspired by QCD \cite{njl}. It has also been applied to the
nucleon, see Refs.~\cite{njln}. However, the lack of confinement makes
the model questionable in the baryonic sector.  It has been shown
that a non-local covariant extension of this model can lead to 
quark confinement for acceptable values of the parameters
\cite{pb}. This occurs due to the fact that the quark propagator has
no real poles, and consequently quarks do not appear as asymptotic
states. There are several other advantages of the non-local version
over the local NJL model: the non-locality regularises the model in a
manner such that anomalies and gauge invariance are
preserved and the momentum-dependent regulator makes the theory finite to
all orders in the $1/N_{c}$ expansion. Finally the dynamical quark
mass is momentum dependent in contrast to the ordinary NJL model and
consistent with lattice simulations of QCD. As a result, the non-local
version of the NJL model  may have more predictive power.

Here we shall use the covariant quark-diquark formalism in a
relativistic Faddeev equation for the nucleon, where we include only
scalar diquark correlations. Due to the separability of the non-local
interaction, the Faddeev equations can be reduced to a set of
effective Bethe-Salpeter equations. This makes it possible to adopt
the numerical method developed for such problems by Oettel \emph{et al.}\ \cite{o}.

\section{The Model }
We consider a non-local NJL model Lagrangian with $SU(2)_{f}\times
SU(3)_{c}$ symmetry. There exists several versions of such non-local
NJL models. Regardless of what version is chosen, a Fierz
transformation allows one to rewrite the interaction in either the
$q\bar{q}$ or $qq$ channels. Here we truncate to the scalar
($0^{+},T=0$) and pseudoscalar ($0^{-},T=1$) mesonic channels. To
investigate the nucleon in the quark-diquark picture, one also needs
to know the $qq$ interaction. We truncate to the
scalar ($0^{+},T=0$) colour $\overline{3}$ $qq$ channel,
\begin{eqnarray}\label{n1}
\mathcal{L}_{I\pi}&=&\frac{1}{2}g_{\pi} j_{\alpha}(x)j_{\alpha}(x),\hspace{2cm}\mathcal{L}_{Is}=g_{s}\overline{J}_{s}(x)J_{s}(x),\nonumber\\
j_{\alpha}(x)&=&\int
d^{4}x_{1}d^{4}x_{3}f(x-x_{3})f(x_{1}-x)\overline{\psi}(x_{1})\Gamma_{\alpha}\psi(x_{3}),\nonumber\\
\overline{J}_{s}(x)&=&\int
d^{4}x_{1}d^{4}x_{3}f(x_{1}-x)f(x-x_{3})\overline{\psi}(x_{1})\big[\gamma_{5}C\tau_{2}\beta^{A}\big]\overline{\psi}^{T}(x_{3}),\nonumber\\
J_{s}(x)&=&\int
d^{4}x_{2}d^{4}x_{4}f(x_{2}-x)f(x-x_{4})\psi^{T}(x_{2})\big[C^{-1}\gamma_{5}\tau_{2}\beta^{A}\big]\psi(x_{4}).\
\end{eqnarray}
Here $\Gamma_{\alpha}=(1,i\gamma_{5}\tau)$ and $m_{c}$ is the current
quark mass of the $u$ and $d$ quarks. The  matrices $\beta^{A}=\sqrt{3/2}
\lambda^{A}(A=2, 5, 7)$ project onto the colour $\overline{3}$
channel with normalisation $\tr (\beta^{A}\beta^{A'})=3\delta^{AA'}$
and the ${\tau_{i}}$'s are flavour $SU(2)$ matrices with $\tr
(\tau_{i}\tau_{j})=2\delta_{ij}$. The object $C=i\gamma_{2}\gamma_{5}$
is the charge conjugation matrix. Since we do not restrict ourselves
to specific choice of interaction, we shall treat the couplings
$g_{s}$ and $g_{\pi}$ as independent.  For simplicity, we assume the
form factor $f(x-x_{i})$ to be local in momentum space, since it  leads to a
separable interaction in momentum space.

The dressed quark propagator $S(k)$ is now constructed by means of
a Schwinger-Dyson equation (SDE) in rainbow-ladder
approximation. Thus the dynamical
constituent quark mass, arising from spontaneously broken chiral
symmetry, is obtained as [the symbol $\Tr$ denotes a trace
over flavour, colour and Dirac indices]
\begin{equation}
m(p)=m_{c}+ig_{\pi}f^{2}(p)\int \frac{d^{4}k}{(2\pi)^{4}} \Tr [S(k)] f^{2}(k),\hspace{1cm}S^{-1}(k)=\not{k}-m(k).\label{n3}
\end{equation}
Following Ref.~\cite{pb}, we choose the form factor to be Gaussian in
Euclidean space, $f(p_{E})=\exp(-p_{E}^{2}/\Lambda^{2})$, where
$\Lambda$ is a cutoff of the theory. If one assumes that $\Lambda$ is
related to the average inverse size of instantons $1/\bar{\rho}$, its
choice thus parametrises non-perturbative properties of the QCD
vacuum.
The choice (\ref{n3}) respects Poincar\'e invariance and leads to quark,
but not colour, confinement, when the dressed quark propagator has no
poles at real $p^{2}$ in Minkowski space. This occurs for
\begin{equation} \label{n4}
\frac{m(0)-m_{c}}{\sqrt{m^{2}_{c}+\Lambda^{2}}-m_{c}} >
\frac{1}{2}\exp\left(-\frac{(\sqrt{m^{2}_{c}+\Lambda^{2}}+m_{c})^{2}}{2\Lambda^{2}}\right).
\end{equation}
For large enough values of the dynamical quark mass, the quark
propagator has no real poles and quarks do not appear as asymptotic
states. The propagator still has infinitely many pairs of complex
poles, both for confining and non-confining parameter sets. This is a
feature of such models and due care should be taken in
handling such poles, which can not be associated with  asymptotic states if the theory is to satisfy unitary.

Our model has four free parameters: the current quark mass $m_{c}$,
the cutoff ($\Lambda$), the coupling constants $g_{\pi}$ and $g_{s}$.
We fix the first three to give a pion mass of $136.6$ MeV with decay
constant of $92.4$ MeV, as the value of the zero-momentum quark mass
in chiral limit $m_{0}(0)$.  We analyse three sets of parameters, as
indicated in table 1. Set $A$ is non-confining and sets $B$ and $C$
are confining (i.e., they satisfy the condition Eq.~(\ref{n4})).
The parameter $g_{s}$ is not constrained by these conditions, which
allows us to analyse the coupling constant dependence through the
ratio $r_{s}=g_{s}/g_{\pi}$.

\begin{table}
\begin{tabular}{rrrr}
\hline
\tablehead{1}{r}{b}{Parameter} &\tablehead{1}{r}{b}{set A} & 
\tablehead{1}{r}{b}{set B} & \tablehead{1}{r}{b}{set C} \\
\hline
$m_{0}(0)$ (MeV) & 250 &350 & 400 \\
$m(0)$ (MeV) & 297.9 &406.2 & 461.3 \\
$m_{c}$ (MeV) & 7.9 &14.4 & 17.6 \\
$\Lambda$ (MeV) & 1046.8 &723.4 & 638.1 \\
$g_{\pi}(\text{ GeV}^{-2})$ &31.6 & 89.0 & 132.6 \\
\hline
\caption{The model parameters for different sets, fitted as discussed in text. The resulting value of the dynamical quark mass $m(0)$ also is shown.}
\end{tabular}
\end{table}

\section{Diquark and Nucleon solution}
In the rainbow-ladder approximation  the diquark $T$-matrix  can be
 written as
\begin{eqnarray}
T(p_{1},p_{2},p_{3},p_{4})&=&f(p_{1})f(p_{3})\big[\gamma_{5}C\tau_{2}\beta^{A}\big]t(q)\big[C^{-1}\gamma_{5}\tau_{2}\beta^{A}\big]f(p_{2})f(p_{4})\nonumber\\
&&\times\delta(p_{1}+p_{2}-p_{3}-p_{4}),\label{d}\
\end{eqnarray}
where $q=p_{1}-p_{3}=p_{4}-p_{2}$ and
\begin{eqnarray}
t(q)&=&\frac{2 g_{s}i}{1+g_{s}J_{s}(q)},\label{n5} \\
 J_{s}(q)&=& i \Tr \int
 \frac{d^{4}k}{(2\pi)^{4}}f^{2}(-k)\big[\gamma_{5}C\tau_{2}\beta^{A}\big]S(-k)^{T}\big[C^{-1}\gamma_{5}\tau_{2}\beta^{A}\big]S(q+k)f^{2}(q+k).\label{n6}\
\end{eqnarray}
In the above equation the quark propagators are the solution of the
rainbow SDE Eq.~(\ref{n3}). The diquark bound state is located at the
pole of the $T$-matrix. The denominator of Eq.~(\ref{n5}) is the same
as that in pionic channel, thus one may conclude that at $r_{s}=1$
the diquark and pion are degenerate. This puts an upper limit to the choice
of $r_{s}$, since diquarks should not to condense in vacuum.

The relativistic Faddeev equation can be written as an effective
two-body BSE between a quark and a diquark due to the separability of
the two-body interaction in momentum-space.  Using the 
the on-shell approximation this becomes 
an eigenvalue problem \cite{o},
\begin{eqnarray}
&&\int \frac{d^{4}k}{(2\pi)^{4}} K_{\alpha\gamma}(p,k;P)\psi_{\gamma\beta}(k,P)
=\lambda(P^{2})\phi_{\alpha\beta}(p,P),\label{n7}\\
&&\psi(p,P)=S(p_{q})D(p_{d})\phi(p,P),\label{n8}\\
&&K(p,k;P)=\frac{2}{g_{dqq}^{2}}X(P_{1})S^{T}(q)\overline{X}(P_{2}).\label{n9}\
\end{eqnarray}
Here $\psi(p,P)$ is the nucleon bound-state wave function related
to the vertex function $\phi(p,P)$ by truncation of the legs. The parameters
$p$ and $P$ are the relative and total momenta in the quark-diquark system,
respectively. We define $p_{q}=\eta P+p$, $p_{d}=(1-\eta)P-p$ and
$q=-p-k+(1-2\eta)P$. The relative momentum of quarks in the
diquark vertices $X$ and $\overline{X}$ are defined as $
p_{1}=p+k/2-(1-3\eta)P/2$ and $p_{2}=-k-p/2+(1-3\eta)P/2$
respectively. (For $\eta$ see below.)

These equations should be solved for the largest eigenvalue of
$\lambda(P^2)$ with the constraint that $ \lambda(P^2)=1/g^{2}_{dqq}$
at $P^2=M^{2}_{N}$, where $M_{N}$ is nucleon mass. Here $g_{dqq}$
denotes the diquark-quark interaction coupling, 
$g_{dqq}^{-2}=\left(\partial_{k^2} J_{s}\right)|_{k^{2}=M_{d}^{2}}$,
where $J_{s}$ is defined in Eq.~(\ref{n6}). In the usual
on-shell approximation the scalar diquark propagator  is taken
to be $D(p_{d})^{-1}=-(p_{d}^{2}+M_{d}^{2})$ where $M_{d}$ is the 
pole of the scalar diquark $T$-matrix. The
functions $X$ and its adjoint $\overline{X}=\gamma_{0}X^{\dag}\gamma_{0}$
are the vertex functions of two quarks within the diquark and can be
obtained from Eq.~(\ref{d}),
\begin{equation}
X(p_{1},k_{d})=g_{dqq}\gamma_{5}C\tau_{2}\beta^{A}f(p_{1}+(1-\sigma)k_{d})f(-p_{1}+\sigma k_{d}).\label{n10}
\end{equation}
The non-locality of our model naturally leads to an extended diquark,
and provides a sufficient regularisation of the ultraviolet
divergences for a diquark-quark loop.  Above we have introduced the
Mandelstam parameters $\eta$ and $\sigma$, $\eta,\sigma \in [0,1]$. In
principle, the eigenvalues should not depend on these parameters if
the formulation is Lorentz covariant. However, in practise due to
numerical errors and singularities in the propagators, one usually only
finds that there are substantial plateaus where the results are
$\eta$- and $\sigma$-independent. The nucleon vertex function
$\phi(p,P)$ should correspond to a state of positive energy, positive
parity and spin-$1/2$. This leads to the general structure
\begin{equation}
\phi(p,P)=\left(S_{1}+\frac{i}{M_{N}}\not{\hspace*{-0.16em}p}S_{2}\right)\Lambda ^{\dag},
\end{equation}
where \(\Lambda ^{\dag}=\frac{1}{2}(1+\frac{\not{P}}{iM_{N}})\) is a
positive energy projection for a nucleon with mass $M_{N}$, and
$S_{1}$ and $S_{2}$ are scalar function. We now solve the
equations in the rest frame of the nucleon. We expand the
scalar functions in terms of Chebyshev polynomials of the first kind,
which turns out to be very efficient for such problems \cite{o}.

\section{Results}
The scalar diquark mass $M_{D}$ is obtained as a pole of
Eq.~(\ref{n5}). We find that for a wide range of $r_{s}$, regardless
of the parameter set and confinement, a bound scalar diquark
exists. The diquark masses for various value of $r_{s}$ for different
parameter sets are plotted in Fig. 1a. 
For sets A, B, C no bound scalar diquark exists for $r_{s}< 0.110,0.112,0.146$,
respectively. Please note that the confinement of the diquark
for very small $r_{s}$ is due to the screening effect of the ultraviolet
cutoff and should not be associated to the underlying QCD confinement
which stems from infrared divergence of gluon and ghost
propagators. Having said that it is possible that  diquark confinement may arise beyond
the rainbow-ladder approximation \cite{new}.  

In order to solve the Faddeev equation numerically, we have first used
the on-shell approximation. We observe that one can not generate a
three-body bound state in this model, without an artificial
enhancement of the quark-diquark coupling of about 3. A similar
feature has been observed in the ordinary NJL model as well
\cite{njln}, but the effect is even more severe here. As can be seen
from Fig.~1 a decrease in $r_{s}$ leads to a larger diquark mass, and
an increase in the off-shell contribution to the $qq$ $T$-matrix. It
is this off-shell contribution that leads to a bound nucleon, and they
are indeed substantial as can be seen in Fig.~1b.  A preliminary
result is shown in Fig.~1c. We also show a fictitious diquark-quark
threshold defined as $M_{qq}+m_{q}^{\text{p}}$, where
$m_{q}^{\text{p}}$ is the real part of the first pole of quark
propagator. Given this definition the diquarks in the nucleon are much more loosely
bound in the non-local model than in the ordinary NJL
model. Nonetheless, we obtain a bound nucleon solution near its
experimental value.

The  research of NRW and MCB was supported by
the UK EPSRC; AHR acknowledges the award of an ORS studentship.
\begin{figure}
\begin{tabular}{ccc}
\includegraphics[height=.15\textheight]{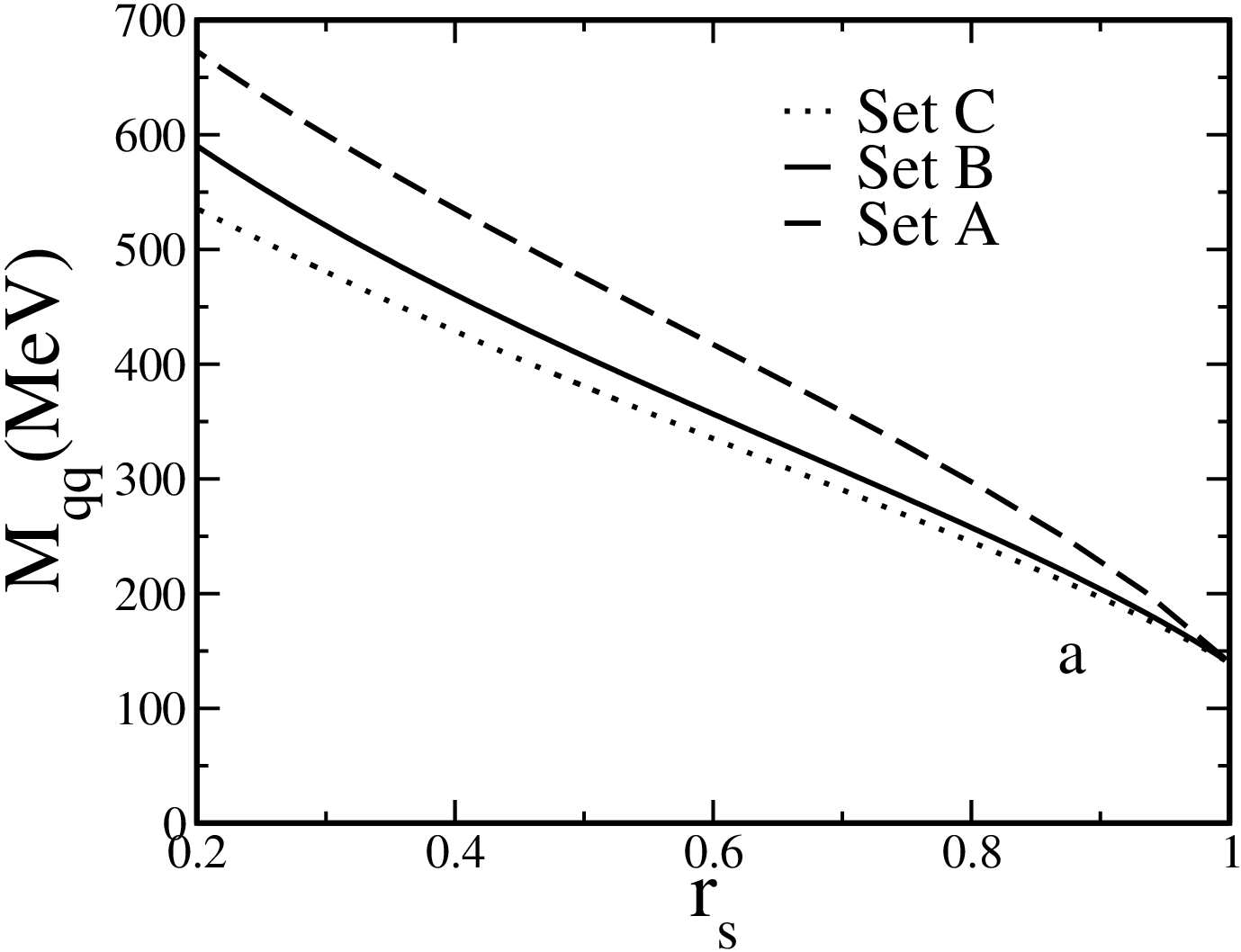}&
\includegraphics[height=.15\textheight]{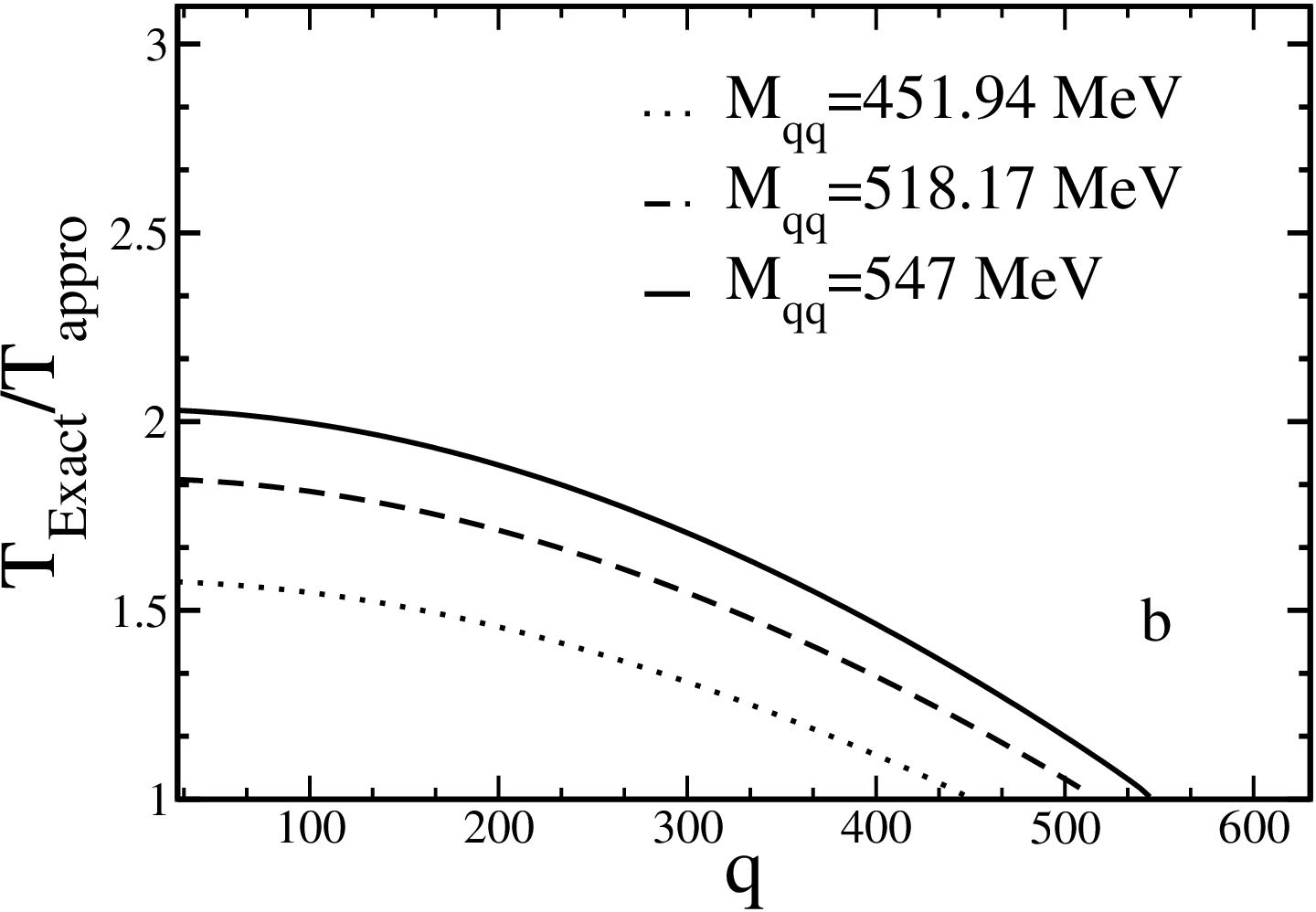}&
\includegraphics[height=.15\textheight]{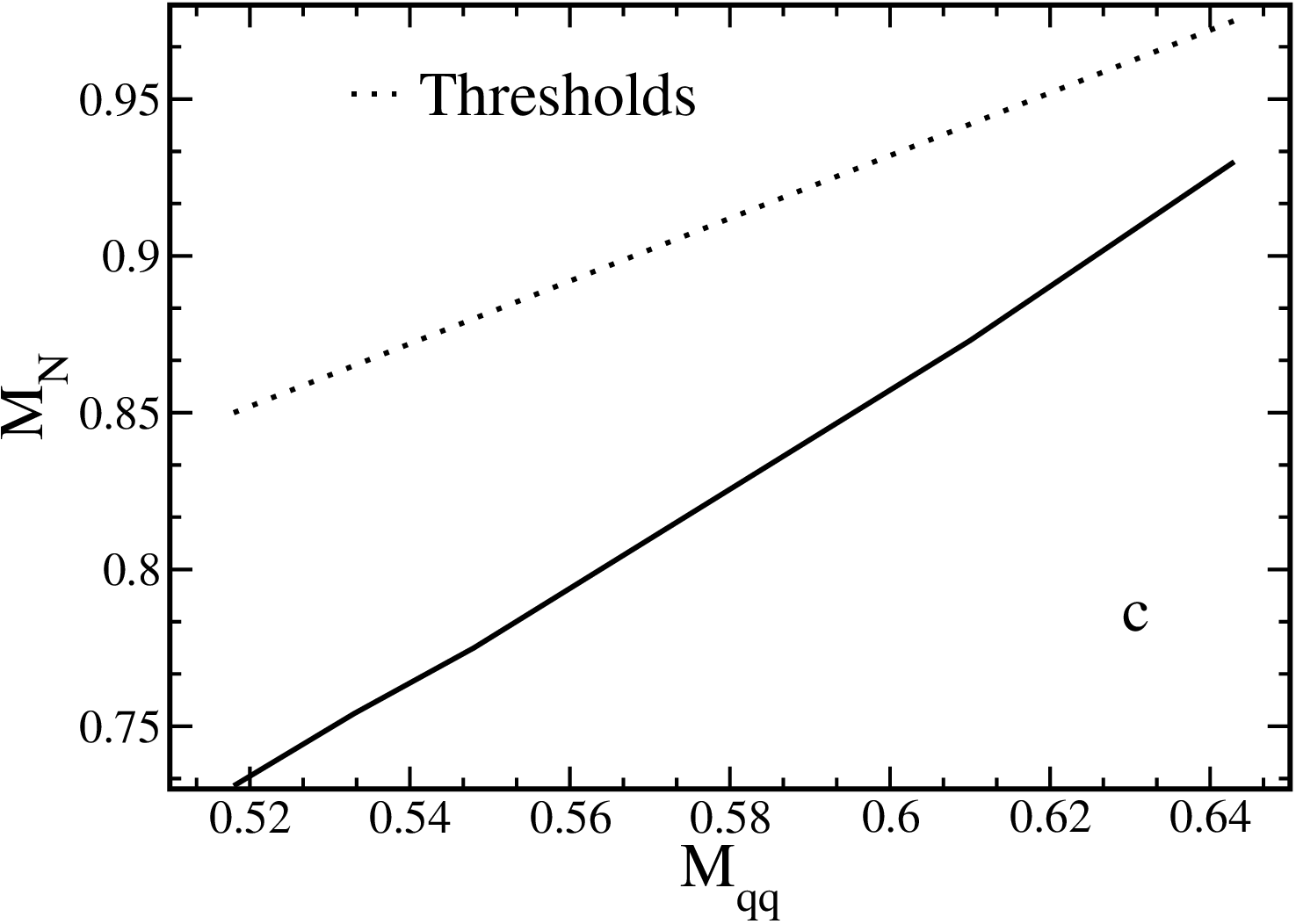}
\end{tabular}
\caption{(a) Shows the diquark masses with respect  to $r_{s}$ for 
three sets of parameters.  (b) Shows the discrepancy between
the on- and off-shell approximation for three diquark masses (parameter set
B). In (c) the  nucleon mass is shown as a function of diquark mass for
set B, the values are given in GeV.}
\end{figure}%

\bibliographystyle{aipproc}

\end{document}